\documentclass[letterpaper,10pt]{article}
\usepackage[margin=1in]{geometry}

\usepackage{graphicx} 
\usepackage{makecell}
\usepackage{bm}
\usepackage{multirow}
\usepackage{amsmath}
\usepackage{mathtools}
\usepackage{makecell}
\usepackage{natbib}
\usepackage{wrapfig}
\usepackage{caption}
\usepackage{algorithm}
\usepackage{algorithmic}
\usepackage[table]{xcolor}
\usepackage[font=small]{caption}

\usepackage{enumitem}
\usepackage[utf8]{inputenc} 
\usepackage[T1]{fontenc}    
\usepackage[hidelinks]{hyperref}       
\usepackage{url}            
\usepackage{booktabs}       
\usepackage{amsfonts}       
\usepackage{nicefrac}       
\usepackage{microtype}      
\usepackage{xcolor}         
\usepackage{xspace}
\newcommand{\ourtitle}{TEE-X\xspace}
\usepackage{pifont}

\newcommand{\imsixtysix}{76.44}

\title{TEE-X: TEE-aware Acceleration
Framework for Large Vision Models at the Edge}

\date{}

\DeclareMathOperator*{\argmax}{arg\,max}

\begin{document}

\title{TEE-X: TEE-aware Acceleration
Framework for Large Vision Models at the Edge}

\date{}

\author{
\small
Kurt M Wilson$^{2,*}$,
Mohaiminul Al Nahian$^{1,*}$,
Abeer Matar A. Almalky$^{1,*}$,
Sadat Shahriyar$^{3,*}$, \\
\small
Souvik Kundu$^{4}$, 
Zhishan Guo$^{2}$,
Abdullah Al Arafat$^{3}$, 
Adnan Siraj Rakin $^{1}$\\
\small
$^{1}$Binghamton University (SUNY) \quad
$^{2}$North Carolina State University \quad
  $^{3}$Florida International University \quad
  $^{4}$Intel\\
  \small $^*$ Equal contribution
}

\maketitle

\begin{abstract}

Despite their remarkable success, machine learning models, particularly in vision applications, are alarmingly vulnerable to a range of security threats. One key factor in the attack landscape is the distinction between white-box and black-box threat models, as the latter poses challenges that limit attack effectiveness when access to model information is limited. As a result, using Trusted Execution Environments (TEEs) enhances security for machine learning applications by protecting model confidentiality and execution integrity, effectively shifting the execution environment from the white-box to the black-box side of the threat model spectrum. While adopting TEEs for large vision models, e.g., Vision Transformers (ViTs), is crucial for enhancing security and privacy, significant challenges related to memory constraints and increased computational latency must be addressed, especially in time-sensitive edge applications where safety and privacy are paramount. The objective of this work is to enable large vision models to be fully hosted within TEEs, achieving GPU-level inference latency for time-sensitive edge vision applications while maintaining performance. To this end, we propose \emph{\textbf{\ourtitle}}, a TEE-aware acceleration framework that introduces a sensitivity-aware modularization technique and enables vectorization in TEE inference. This design is validated on OP-TEE for Arm TrustZone, configured to optimize performance on the NVIDIA Jetson AGX Xavier for efficient edge vision applications using ViT models. The findings reveal that \ourtitle delivers an effective TEE-aware acceleration framework that achieves minimal accuracy-latency trade-offs while ensuring fast and secure edge inference for vision models.

\end{abstract}
\section{Introduction}

\begin{wrapfigure}{r}{0.45\textwidth}
\centering
\vspace{-15pt}
\includegraphics[width=0.9\linewidth]{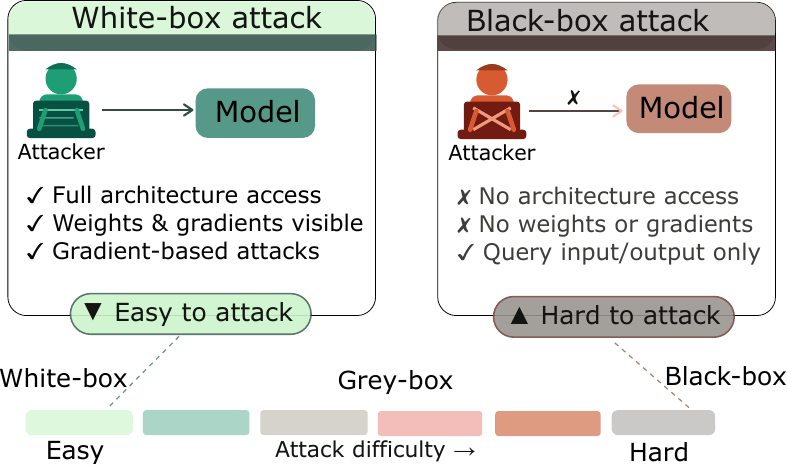}

\caption{\em Overview of attack types categorized according to different threat model settings.}
\label{fig:overview}
\vspace{-12pt}
\end{wrapfigure}
Machine learning (ML) models have made significant strides across various tasks, demonstrating strong performance in many areas, particularly in vision applications~\cite{chen2021crossvit,carion2020end}.  Despite their success, prior studies have demonstrated that these models remain highly vulnerable to a broad spectrum of security attacks, including: memory fault injections~\cite{yao2020deephammer, dong2023one,rakin2019bit,rakin2020tbt,wang2025your,ahmed2024deep,zheng2023trojvit,zhou2024makes}, side-channels~\cite{kim2014flipping,hong2020securityanalysisdeepneural,gruss2016rowhammer,seaborn2015exploiting,van2016drammer,197231,zhang2020pthammer,lin2025gpuhammer,lin2026gpubreach}, and adversarial attacks~\cite{carlini2017towards,liu2016delving,zhao2023evaluating,ilyas2019adversarial,akhtar2018threat,10.1145/3052973.3053009,mahmood2021robustness,joshi2021adversarial,wang2022generating,yuan2023you,naseer2021improving,han2022enhancing,lv2021dbia}. These vulnerabilities highlight the ongoing need for improved security measures in the deployment of ML models, especially in vision applications, that are susceptible to these attacks.

As shown in Figure~\ref{fig:overview}, these attacks are generally classified based on their threat assumptions, which range from white-box to black-box scenarios. In white-box attacks, adversaries are assumed to have complete access to the target model, including its architecture and parameters~\cite{akhtar2018threat,mahmood2021robustness,joshi2021adversarial,wang2022generating,yuan2023you,naseer2021improving,han2022enhancing,lv2021dbia}, thereby substantially simplifying the attack process and often leading to a significant impact on model security. In contrast, black-box attacks represent a considerably more challenging threat model, where adversaries have no direct access to the model internals and can only interact with the system through input-output queries or indirect observations~\cite{wang2022generating,shi2022decision,tang2024black}. Consequently, the effectiveness and feasibility of attacks in black-box settings are substantially constrained compared to those in white-box settings.

As a result, being on the black-box side of the spectrum has motivated the security community to adopt Trusted Execution Environments (TEEs) for ML applications, which provide an isolated and protected execution enclave often constraining the attack threat model more closer to a black-box spectrum~\cite{hou2021model,mo2020darknetz,sun2020shadownet,lee2019occlumency,hanzlik2021mlcapsule,li2021lasagna,shen2020occlum,shen2022soter,tramer2018slalom,nayan2025secureinfer}. These TEEs have been adopted as a defensive mechanism to enforce two primary objectives, model confidentiality (privacy) and execution integrity during inference (security). By isolating model execution within protected hardware boundaries, TEEs effectively shift the attack surface from the white-box to the black-box side of the spectrum (as shown in Fig.~\ref{fig:overview}). While adopting TEEs for ML models to provide security and privacy protection is pivotal, the key challenge in adopting TEEs, especially for large vision models such as Vision Transformers (ViTs), stems from the memory and computational (e.g., latency) bottlenecks they introduce. Deploying a large vision model inside a TEE presents three primary bottlenecks: i) The memory capacity of the TEE is limited, which makes it challenging for some of the larger models to fit inside a TEE. ii) Executing these models within TEE introduces considerable computational overhead and inference latency compared to native GPU execution. iii) The above two challenges become further exacerbated in edge applications~\cite {xu2025edge}, where inference is time sensitive, at the same time, TEE-enabled inference is even more important in edge applications where safety (e.g., self-driving car~\cite{prakash2021multi}) and privacy( e.g, healthcare~\cite{javaid2024computer}) are a priority.

To address this, existing approaches for deploying ML models within TEEs have adopted two alternative design choices. The first category is based on model partitioning, where parts or layers of the network are executed within the secure enclave, while the remaining components are offloaded to the untrusted environment to alleviate the stringent memory constraints imposed by TEEs~\cite{mo2020darknetz,sun2020shadownet,lee2019occlumency,tramer2018slalom}. The second category adopts operation-aware execution strategies, selectively placing sensitive operations inside the TEE while offloading computationally intensive operations to on-chip memory or external accelerators to improve execution efficiency~\cite{hou2021model,nayan2025secureinfer}. While these methods are effective at partially reducing inference costs (e.g., latency), they still fall short of GPU-only inference time, which is critical, especially for edge applications. At the same time, exiting the TEE at any stage of model inference still leaves the attacker with additional information (e.g., layers) outside the TEE to exploit.  This motivated our proposed approach to designing an \textit{acceleration method that enables hosting entire large vision models inside TEEs while achieving on-par GPU inference latency and maintaining model performance}.

\begin{wrapfigure}{r}{0.4\textwidth}
\centering
\vspace{-10pt}
\includegraphics[width=\linewidth]{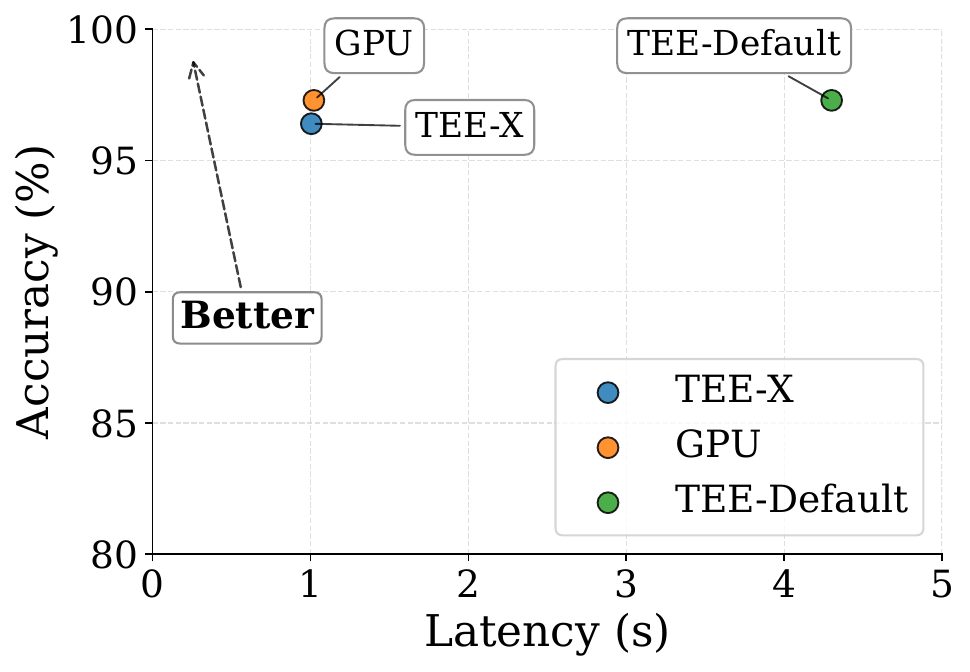}
\vspace{-4mm}
\caption{\em Comparison of DeiT-small inference in: GPU, TEE, and accelerated using \ourtitle.}
\label{fig:GPU_TEE_compare}
\vspace{-8pt}
\end{wrapfigure}

To this end, we propose \textbf{\ourtitle}, a TEE-aware acceleration framework to run the entire model inference inside secure enclaves. Unlike prior approaches that rely on model partitioning or offloading operations to untrusted accelerators, \ourtitle enables complete in-enclave execution by designing a novel sensitivity-aware modularization technique combined with TEE-aware computational budget assignment for each module. On the TEE side, for the first time, vectorization is enabled using Arm Neon SIMD instructions to perform a multiply-accumulate operation, providing GPU-like parallelism. The proposed optimization is designed to achieve two objectives: i) \ourtitle ensures secure, fully protected model inference using the strong security benefits of TEEs. ii) Meeting the performance demands of large-scale edge vision applications (ref. Figure~\ref{fig:GPU_TEE_compare}). To evaluate our proposed solution, we stress test \ourtitle on an edge vision application where the evaluation model is ViTs due to their substantially larger model sizes and higher computational complexity compared to other vision architectures, and on TrustZone within an Armv8 edge device (NVIDIA Jetson AGX Xavier) using a limited trusted memory carveout size. The results demonstrate that \ourtitle provides an effective TEE-aware acceleration framework that achieves negligible accuracy--latency trade-offs while ensuring faster, more secure edge inference for vision models.

\section{Background and Related Work}

\textbf{Trusted Execution Environments (TEEs)} are hardware-isolated execution environments designed to securely store sensitive data and models, and to protect computations from external access. Popular TEE technologies include Intel SGX~\cite{mckeen2013innovative}, AMD SEV~\cite{kaplan2016amd}, and Arm TrustZone~\cite{alves2004trustzone}. Despite their security advantages, TEEs suffer from limited memory capacity and performance overhead, which pose challenges for deploying large ML models. Following prior studies~\cite{hou2021model,mo2020darknetz,shen2022model,sun2020shadownet}, we assume that the TEE operates as a trusted enclave within a potentially untrusted host system, including GPUs. Under this threat model, all data, model parameters, and computations inside the TEE are considered secure. Although side-channel attacks against TEEs have been explored in prior works~\cite{zhang2023interface,huo2020bluethunder,van2018foreshadow}, they are outside the scope of this paper.

\textbf{Vision Models inside TEEs.}
Several studies~\cite{hou2021model,mo2020darknetz,sun2020shadownet,lee2019occlumency,hanzlik2021mlcapsule,li2021lasagna,shen2020occlum,shen2022soter,tramer2018slalom,nayan2025secureinfer} have investigated the deployment of vision models within secure enclaves. However, these frameworks largely overlook the unique architectural properties and system-level challenges associated with ViTs. In addition, many existing approaches only partially execute model components inside the enclave, rather than fully encapsulating the entire model within the secure environment. This fragmented execution design also introduces significant and often unavoidable latency overhead compared to standard GPU-based inference. Therefore, this work focuses on developing an efficient acceleration technique that enables full model execution within TEEs while minimizing latency overhead introduced by secure-enclave constraints and closely matching native GPU inference performance.

\section{\ourtitle: TEE-aware Acceleration Framework}

We propose \ourtitle, designed to host large vision models such as ViTs within TEEs, informed by TEEs' strict memory and computation budgets. At the same time, the proposed acceleration should preserve the model performance (i.e., accuracy and latency demand of the edge application). The proposed acceleration mechanism is motivated by two design principles: First, for a given application and TEE, we are bound by strict memory and latency budgets, especially for real-time edge models that must deliver model output within a strict latency window. Second, while maintaining the budget, the model acceleration method must maintain its accuracy. These two design principles are inversely proportional; i.e., in general, a smaller, faster model tends to be less accurate than its larger and higher-latency counterpart~\cite{howard2017mobilenets}. To address this, our proposed \ourtitle has two components: \emph{first}, \emph{\textbf{Modularization}}, which is a TEE-aware computation budget allocation and model optimization technique \& \emph{second},  \emph{\textbf{Vectorization}}, which utilizes Arm Neon vectorization support with fused multiply-accumulation operation and cache-friendly access, in order to boost the Modularization technique in achieving on-par edge GPU performance.

\subsection{Modularization: TEE-aware computation budget allocation}
\label{sec:proposed_modularization_TEE_X}
\begin{figure}[t]
    \centering
    \includegraphics[width=0.8\textwidth]{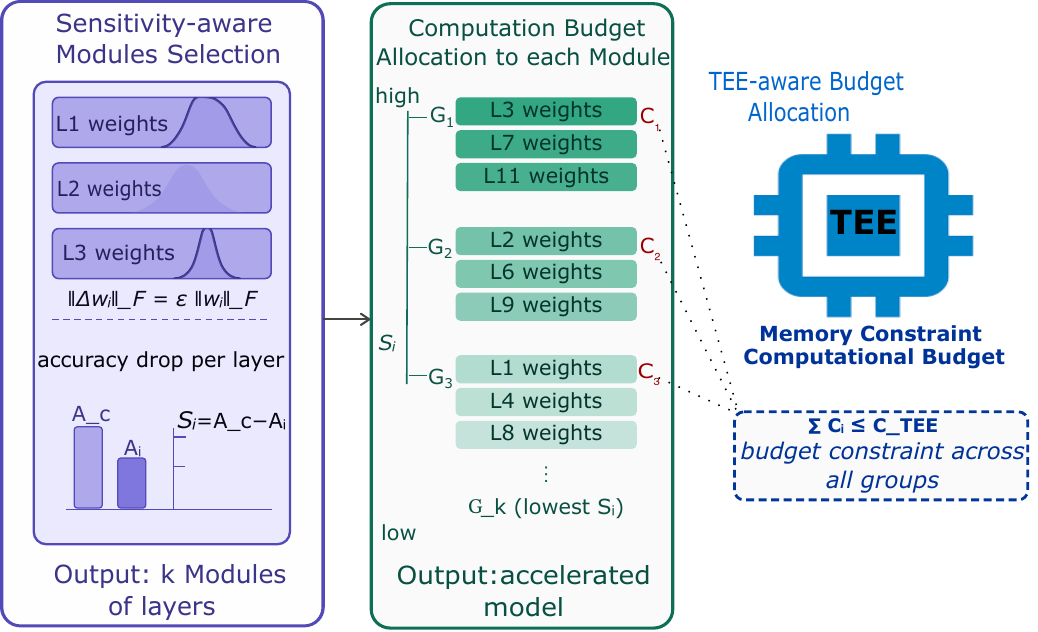}
    \caption{\em Overview of \ourtitle' modularization steps.}
    \vspace{-3mm}
    \label{fig:method}
\end{figure}

The first component modularization technique consists of the following steps: the first component performs a \emph{sensitivity-aware module selection of the weight matrices} of a ViT model that will run inside a TEE with an equal computation budget, where each module consists of a subset of the weight matrices. The intuition behind this design choice is that layers with equal sensitivity should receive a similar computational budget. Once the modules are created, TEE reports the total available computation budget to match GPU acceleration. Informed by this budget, we propose allocating a \emph{computation budget} to each module based on its prior sensitivity ranking, ensuring that the total computation budget is met. Once the module is created and its corresponding budget is allocated, the final step practically \emph{realizes the TEE-aware budget allocation} by learning a transformation of each module's weight vector that reflects the allocated computational budget while performing layer-wise input and weight multiplication. This ensures that allocating a higher computation budget to more sensitive layers preserves the accuracy budget, while an aggressive reduction in the computation budget in less sensitive regions helps maintain the latency budget. The overview of TEE-X is illustrated in Figure~\ref{fig:method}.

\noindent\paragraph{Step-1: Sensitivity-aware Module Selection.}
\label{sec:layer_sen_based_grouping}
Let $W = \{w_1, w_2, \ldots, w_L\}$ be the weights of a pre-trained ViT model consisting of $L$ layers. The goal of this step is to partition the weight matrix into groups with equal sensitivity levels. Hence, we propose to rank the layers based on a layer-wise sensitivity metric $S_i$ (defined in section~\ref{sec:sen_metrics_and_training}). The layers are divided into $K$ modules $\{\mathcal{G}_{k}| k=1,2, \ldots, K\}$, where layers with a similar level of sensitivity are grouped together in the same module.

\noindent\textbf{Step-2: Computation Budget Allocation to Each Module.} \label{Method:part2}
Let $x_i \in \mathbb{R}^{N\times n_{i}}$ be the input of the i-th layer with $w_i \in \mathbb{R}^{n_{i}\times n_o}$. Then the activation/output of the layer is  $x_{i+1} = g(x_i\times w_i)$, where $g(.)$ is an activation function. The key computational demands come from  $x_i\times w_i$. To optimize/minimize the computational load within TEE, we propose to learn a transformation of each $w_i$ such that the overall computation for $x_i\times w_i$ is minimized. Let $w_i$ is transformed to $A_i \in \mathbb{R}^{n_i\times h_i}$ and $B_i \in \mathbb{R}^{h_i\times n_o}$. Now, $h_i$ will control the computation requirements for $x_i\times w_i = (x_i\times A_i)\times B_i$, which we define as the computational budget of the i-th layer.

To speed up the computation within TEE, $h_i$ should be smaller than a factor $\mathcal{C} \propto \frac{T_{gpu}}{T_{TEE}}$, where $T_{gpu}$ and $T_{TEE}$ are per-operation times of GPU and TEE execution, to keep the TEE inference time on par with GPU's. Let $h_k$ be the computational budget of each layer in module $\mathcal{G}_k$ and $n_{g_k}$ be the number of layers in $\mathcal{G}_k$. 
Then the budget assignment to each module can be found by solving:
\begin{equation}
\begin{aligned}
\argmax_{\{h_k\}_{k=1}^{K}} \mathcal{A}(\{h_k\}_{k=1}^{K})
\quad \text{s.t.} \quad
 \sum_{k=1}^{K} n_{g_k} h_k \leq \mathbb{C},
 ~h_k \in \mathbb{Z}^{+}
\end{aligned}
\label{eq:budget_opt}
\end{equation}
here, $\mathcal{A}(\{h_k\}_{k=1}^{K})$ denotes the model accuracy obtained after assigning dimension $h_k$ to module $\mathcal{G}_k$ and $\mathbb{C}$ is available computational budget of TEE.

\noindent\textbf{Step-3. Realization of TEE-aware Budget Allocation.} \label{method:part1}
Let $\mathcal{T}:w_i \to A_i$ be a learned transformer and $B_i: A_i \to w_i$ is a linear transformer such that $w_i = A_i \times B_i$. Once the computational budgets from Step-2 are found, all such $\mathcal{T}$'s and $B_i$'s can be learned to decompose each weight matrix in the model. Note that solving Eq.~\ref{eq:budget_opt} exactly is computationally expensive. 
The objective function is not available in closed form and depends on training $\mathcal{T}, B_i$ for each possible budget configuration $h_k$ for each module subject to the budget constraint, $\mathbb{C}$. 

Therefore, instead of solving Eq.~\ref{eq:budget_opt} directly, we use a sensitivity-guided rule-based allocation. We divide this total computation budget among modules based on the aggregate sensitivity score of a module's layers relative to the total sensitivity score of all layers. If the k-th module $\mathcal{G}_k$ has $n_{gk}$ layers, then the assigned compute budget to each layer of that module, $h_k$ is given by
\begin{equation}
    h_k=\frac{{\sum_{\forall j \in \mathcal{G}_k}}S_{j}} { \sum_{\forall i \in L} S_i } \cdot \frac{\mathbb{C}}{n_{gk}}
    \label{eq:rank_assignment}
\end{equation}
Eq.~\ref{eq:rank_assignment} ensures that we remain within computation budget, i.e., $\sum{n_{gk}\cdot h_k}= \mathbb{C}$. Once we have assigned this budget to each module, we jointly train the transformation networks across all modules.

\subsection{Proposed Sensitivity Metric and Training the Transformation Networks}
\label{sec:sen_metrics_and_training}

\noindent\textbf{Proposed Layer Sensitivity Metric ($S_i$).}
To measure the sensitivity of each layer, we perturb one layer at a time while keeping the others fixed. For the $i$-th layer, we sample a random Gaussian noise tensor
\(
g_i \sim \mathcal{N}(0, I),
\)
where $g_i$ has the same dimension as $w_i$. 
The weight perturbation \(\Delta w_i\) is defined as: 
\[
\Delta w_i =
\epsilon \|w_i\|_F \cdot
\frac{g_i}{\|g_i\|_F},
\]
where $\epsilon$ is the perturbation budget such that the Frobenius norm of the weight perturbation $\|\Delta w_i\|_F$ is proportional to the Frobenius norm of the layer weights, $\|w_i\|_F$.

The perturbed layer weight is given by $\widetilde{w}_i = w_i + \Delta w_i$.
Only $w_i$ is replaced by $\widetilde{w}_i$ during the evaluation, while the remaining layers are kept unchanged.

Let $\mathcal{A}_{\mathrm{w}}$ denote the accuracy of the original model, and let $\mathcal{A}_{\widetilde{w}_i}$ denote the accuracy after perturbing the $i$-th layer. The sensitivity score, $S_i$ of layer $i$ is defined as the resulting accuracy drop:
\begin{equation}
   S_i = \mathcal{A}_{\mathrm{w}} - \mathcal{A}_{\widetilde{w}_i}  
\end{equation}

A larger $S_i$ indicates that the layer is more sensitive to perturbation, which is used as a criterion for our TEE computation budget allocation. 

\noindent\textbf{Training of the Transformation Network.} 
To realize the acceleration framework proposed in Section~\ref{sec:proposed_modularization_TEE_X}, we need to train the transformation networks $\mathcal{T}_k$ and $B_k$ for each of the $\mathcal{G}_k$ modules.
Let $F(\cdot)$ denote the original pretrained ViT model, and let $\hat{F}(\cdot)$ denote the transformed model consisting of the transformation networks $\mathcal{T}_k$ and $B_k$.
For an input sample $x$ with label $y$, the transformed model produces $\hat{y}=\hat{F}(x).$ We optimize the transformed model using three complementary losses, defined as follows:
\begin{align*}
\mathcal{L}_{\mathrm{CE}}
&=
-\sum_i y_i \log(\hat{y}_i), \quad
\mathcal{L}_{\mathrm{KD}}
 =
\tau^2
\mathrm{KL}
\left(
\sigma\left(F(x)/\tau\right)
\;\middle\|\;
\sigma\left(\hat{F}(x)/\tau\right)
\right), \\
&\mathcal{L}_{\mathrm{MSE}}
=
\frac{1}{L}
\sum_{i=1}^{L}
\left\|
x_i w_i - \hat{x}_{i+1}
\right\|_2^2 .
\end{align*}
here, $\sigma(\cdot)$ denotes the softmax function, and $\tau$ is the distillation temperature. 
The term $x_i w_i$ represents the original output of the $i$-th layer, while $\hat{x}_{i+1}$ represents the corresponding output produced by the transformed operation.
The cross-entropy loss, $\mathcal{L}_{\mathrm{CE}}$ preserves task performance with respect to the ground-truth labels. 
The knowledge distillation loss, $\mathcal{L}_{\mathrm{KD}}$ preserves the output behavior of the original pretrained ViT. 
The MSE loss, $\mathcal{L}_{\mathrm{MSE}}$ aligns the transformed layer outputs with the original layer outputs. 

All module-wise transformation networks and linear transforms are optimized jointly by solving
\begin{equation}
\min_{\{\mathcal{T}_k,B_k\}_{k=1}^{K}} \mathcal{L}_{\mathrm{total}}
=
 \min_{\{\mathcal{T}_k,B_k\}_{k=1}^{K}} (\mathcal{L}_{\mathrm{CE}}
+
\lambda_{\mathrm{KD}}\mathcal{L}_{\mathrm{KD}}
+
\lambda_{\mathrm{MSE}}\mathcal{L}_{\mathrm{MSE}}),
\end{equation}
where $\lambda_{\mathrm{KD}}$ and $\lambda_{\mathrm{MSE}}$ control the contributions of the distillation and output-alignment losses, respectively. After training, each learned transformation network $\mathcal{T}_k$ is used to generate the transformed weights $A_{k}=\mathcal{T}_k(w_{k})$ for all layers in module $\mathcal{G}_k$. 
During inference, we store only the generated transformed weights $\{A_{k}\}$ and the corresponding module-wise linear transforms $\{B_k\}$,
which reduces both the memory footprint and the computation required inside the TEE.

\noindent\textbf{Computational Budget of TEE ($\mathbb{C}$).} 
To keep the model inference time on par with the GPU inference, we need to compute the maximum TEE computational budget $\mathbb{C}$. To perform $x_i \times w_i$, the GPU inference time is $2Nn_in_oT_{gpu}$, which is an approximation of the amount of FLOP counts times per-operation time. Then the corresponding TEE computational budget $h_i$ for on-par GPU performance is given by:
\begin{equation}
    h_i  = \frac{n_in_o}{n_i + n_o}\cdot \frac{T_{gpu}}{T_{TEE}}
\end{equation}
$$$$
Then, the total computational budget of TEE is $\mathbb{C} = \sum_{\forall i} h_i$  to maintain GPU performance. Our evaluation shows that using this theoretical budget for module-level computation allocation also matches GPU-level performance in practical TEE implementations.

\subsection{Vectorization: TEE Acceleration Support}

\begin{wrapfigure}{r}{0.5\textwidth}
    \vspace{-20pt}
    \centering
    \includegraphics[width=\linewidth, trim=0.2in 0 0 0, clip]{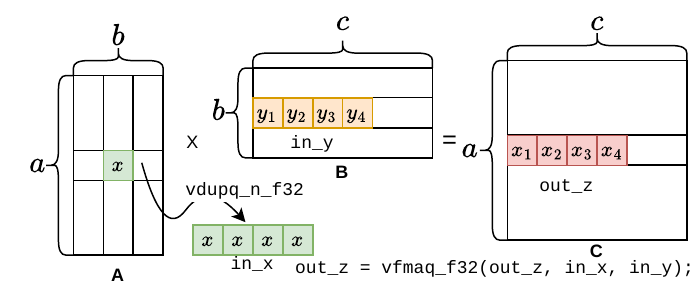}
    \vspace{-8mm}
    \caption{\em Overview of Vectorization Step: Vectorizing $C = A \times B$. }
    \vspace{-20pt}
    \label{fig:matmul_vect}
\end{wrapfigure}

To efficiently perform matrix multiplication and residual addition in the trusted application (TA) within the TEE without a GPU, we use SIMD vectorization wherever possible. We use Arm Neon's 128-bit-wide registers, which can store and operate on four 32-bit floats simultaneously. Combined with Arm's fused-multiply-accumulate instruction and cache-friendly access ordering, we can accelerate tensor operations compared to naive serialized implementations. 
Furthermore, to reduce data dependencies and make efficient use of the instruction pipeline, we unroll the inner loop to write results to four wide registers at a time.

For single batch inference, this allows \ourtitle to match the host/GPU performance for some configurations. For batch sizes beyond 1, which make better use of the parallelism available within the GPU, the performance gap will increase, as the GPU will be able to progress on multiple inputs within the batch at the same time, whereas our parallelism is column-wise within a single input.

\textbf{Illustrative Example. }Figure \ref{fig:matmul_vect} shows an example of a vectorized matrix multiply operation, where we perform the multiply-accumulate step to multiple values within a row. To perform parallel multiply-accumulate operations, we utilize 128-bit wide registers to process four float32 values simultaneously. A scalar from $A$ is broadcast into all four slots of register in\_x (vdupq\_n\_f32), while a 4-float segment from a row in $B$ is loaded into in\_y (vld1q\_f32). These are then processed alongside the current values loaded into out\_z via Fused Multiply-Accumulate (vfmaq\_f32). This allows us to compute four results in parallel before storing out\_z back into the result matrix $C$, significantly increasing computational density. This process uses three 128-bit wide registers.

\begin{wrapfigure}{r}{0.48\textwidth}
    \vspace{-20pt}
    \centering
    \includegraphics[width=\linewidth, trim=0.2in 0 0.3in 0, clip]{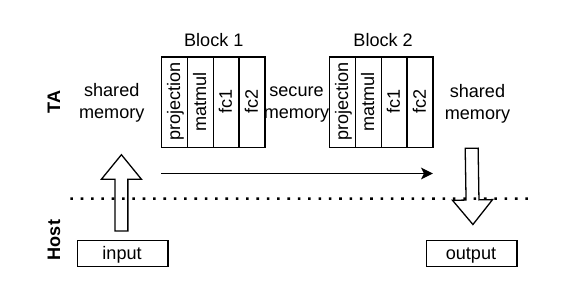}
    \vspace{-7mm}
    \caption{\em Inference of two consecutive blocks in the TA. The input gets sent to the TA as a pointer to shared memory. Working memory and block results are kept in secure memory, except for the outputs of the last block, which gets placed into shared memory. The shared memory pointer gets passed to the host as the output, which can read the result.}
    \vspace{-20pt}
    \label{fig:memory_workflow}
\end{wrapfigure}

\section{Experimental Evaluation}
\label{sec:Experiments}

\subsection{TEE and Hardware implementation details} 
\label{sec:TEE_hardware_implementation}
We implement \ourtitle with OP-TEE on Arm TrustZone. 
The implementation is split into two parts: the host application and the trusted application (TA). The host application runs in a non-secure world and has access to hardware peripherals, storage, and the network. The host has access to all system memory, except for a small TrustZone memory carve-out available only to the trusted application running within TrustZone.

\textbf{Vectorization. }Attempting to use the GPU within the TA could result in exposure of the network weights and intermediary data, so we perform the layer operations on the CPU. To improve CPU inference speed, we use Arm Neon SIMD instructions to perform parallel multiply-accumulate operations, providing some GPU-like parallelism.

During the inner loop of matrix multiplication, we broadcast a single scalar from one operand across all four lanes of a 128-bit float32x4\_t register, and load four contiguous floats from the other operand into a separate vector register. The fused multiply-accumulate instruction (vfmaq\_f32) then produces four results in parallel. To take advantage of pipelining, we unroll across four independent accumulator registers, advancing by 16 columns per inner loop, breaking the dependency chain between successive multiply-accumulate operations. We do similar steps for residuals, bias addition, and layer normalization to update multiple values per iteration.

To use Neon within the TA, OP-TEE must be compiled with CFG\_WITH\_VFP, which adds support for saving and loading Neon registers during context switches. 
This adds two small sources of overhead: For the very first vector instruction in the TA after a context switch, an interrupt records the usage of the Neon registers. On the next context switch, OP-TEE checks whether a vector instruction has run, and if so, stores the state of the extra registers. The overhead is minor compared to the performance gains from vectorization, but it is important to minimize frequent host communication.

\textbf{Transferring Weights and Inputs. }For efficient transfer of encoded weights and inputs to the TA, and to reduce TA memory usage, we used shared memory allocated by the host. The host application loads the weights into normal world memory, and marks it as accessible to the TA with TEEC\_RegisterSharedMemory, and passes it to TA commands as a pointer. Shared memory does not count against the TrustZone memory limit. We minimize communication between the host and TA as much as possible while still balancing memory usage. For example, when consecutive layers are run on the TA, we do not send the intermediate results back and forth to the host. All compressed weights stay loaded in TA memory, and intermediate tensors are always kept in secure memory. We allocate intermediate tensors as soon as they are needed, and deallocate them as soon as they're read for the last time. This reduces peak memory usage, allowing TEE-X to be used on TEEs with smaller memory limits.

\textbf{Hardware Platform. }We evaluate on an Nvidia Jetson AGX Xavier, which has an 8-core 64-bit Armv8.2 CPU at 2.2 GHz and an integrated Volta GPU. We build upon the default Linux filesystem image from Nvidia with OP-TEE. We fix the CPU frequency to 2.2 GHz, and assign the inference process to one CPU core.

\subsection{Evaluation Setting}

\textbf{Models and Datasets.} We evaluate \ourtitle on multiple models: DeiT-Small and DeiT-Base~\cite{touvron2021training}, which are widely used representative ViT architectures with different capacity levels. For benchmarking, we use two standard image classification datasets: CIFAR-10~\citep{krizhevsky2009learning}, consisting of 32$\times$32 images across 10 classes, and ImageNet~\citep{deng2009imagenet}, which contains 224$\times$224 high-resolution images spanning 1,000 classes.

\textbf{Hyperparameters and Experimental Setup.}
\label{sec:hyperparameter}
In all cases, we divide our model into $K=3$ modules. The number of layers per module is fixed to be $(L/3-1), L/3, (L/3+1)$. For DeiT-small and DeiT-base, this translates to $3,4,~\text{and}~5$ layers in each module. For each module, we train transformation networks for $attn\_qkv,~attn\_proj,~mlp\_fc1,~mlp\_fc2$. The network $\mathcal{T}_k$ is a 2 linear layer MLP with pre- and post-affine transform, and $B_k$ is a single linear layer. We exclude the first, last and LayerNorm layers from transformations. The transformation network parameters are quantized to 4-bit with quantization aware training and first and last layer is quantized in 8-bit. For CIFAR-10, we train the models for 500 epochs with learning rate $1e-4$ and $5e-5$ respectively for small and base models with batch size 256. For ImageNet, we train for 300 epochs with a batch size 1024 and a learning rate $2e-4$. We set $\lambda_{\mathrm{KD}}$ to be $1e1$ and $1e3$ for CIFAR-10 and ImageNet respectively and $\lambda_{\mathrm{MSE}}$ is chosen to be $1.0$. Our most demanding experiment can be performed using a single $A6000$ GPU with $48\text{GB}$ VRAM.

\textbf{Evaluation Criteria}
We evaluate \ourtitle to answer four key criteria: \textbf{(C1)} preserving the ViT accuracy, \textbf{(C2)} achieving GPU-comparable inference latency, \textbf{(C3)} outperforming existing acceleration techniques and TEE-based frameworks, and \textbf{(C4)} improving security by transitioning models toward a black-box setting.

\textbf{\ourtitle Accuracy and Latency (\textit{Answers for C1 and C2}) Trade-offs.} 
Table~\ref{tab:DeiT_small} demonstrates that \ourtitle achieves substantial model size reduction while preserving high classification accuracy across different computation reduction factors $\mathcal{C}$. Under $\mathcal{C}=0.5$ and $\mathcal{C}=0.4$, the accuracy degradation remains below 1\% while reducing the model size by more than an order of magnitude. Even with the aggressive reduction setting of $\mathcal{C}=0.2$, where the DeiT-Small model is compressed by approximately $47\times$, the accuracy drop is limited to only 2.65\%.

In terms of inference latency, \ourtitle provides a flexible trade-off between accuracy and execution speed. For a higher computation budget ( $\mathcal{C}=0.5$), \ourtitle preserves near-baseline accuracy with comparable GPU inference latency, while more aggressive settings ( $\mathcal{C}=0.2$) further reduce inference time and can even outperform GPU-based execution using proposed modularization and vectorization support. These results demonstrate that \ourtitle outperforms existing compact vision models while enabling efficient and secure execution within TEEs.

Overall, these results satisfy \textit{C1} and \textit{C2} by showing that \ourtitle effectively reduces the memory and computational overhead of ViTs, enabling full deployment inside edge TEEs while maintaining strong accuracy and efficient inference performance.

\begin{table}[t]
\centering
\vspace{-15pt}
\caption{Model size, Accuracy, inference time, and memory of DeiT-Small (Original vs Inside TEE) on CIFAR-10 under different computation reduction factors $\mathcal{C}$. Accuracy (\%) is reported. Peak Mem. (MB) is the peak memory utilization inside a TEE at inference time.}
\label{tab:DeiT_small}
\scriptsize
\setlength{\tabcolsep}{4pt}
\begin{tabular}{lccccc}
\toprule
\textbf{Method} & \textbf{Model Size (MB)} & \textbf{Accuracy} & \textbf{Avg Layer Time (ms)} & \textbf{Avg Time (ms)} &\textbf{Peak Mem. (MB)} \\
\midrule
Baseline Host (GPU) & 82.66 & 97.29  & 82.62 & 1022.35 & 154.78\\
\ourtitle~$(\mathcal{C}=0.5)$      & 3.48 (23$\times$) & 96.48 & 58.24 (1.42$\times$) & 1082.67  & 21.91 \\
\ourtitle~$(\mathcal{C}=0.4)$     & 2.93 (28$\times$)  & 96.40 & 51.76 (1.60$\times$) & 1006.74 &18.14\\
\ourtitle~$(\mathcal{C}=0.2)$     & 1.75 (47$\times$) & 94.64 &  48.02 (1.72$\times$)&  957.13 & 10.13\\
\bottomrule
\end{tabular}
\vspace{-15pt}
\end{table}

\textbf{Comparison with Competitive Methods (\textit{Answers for C3}).} We evaluate \ourtitle from two complementary perspectives: SOTA TEE-based frameworks and traditional model acceleration techniques.

\begin{wraptable}{r}{0.4\textwidth}
\centering
\vspace{-10pt}
\caption{Impact of \ourtitle's modularization compared to SOTA TEE-based model partitioning. Since this is a software optimization comparison, we applied vectorization acceleration to the competition as well.}
\scriptsize
\label{tab:tee_comparison}
\setlength{\tabcolsep}{5pt}
\renewcommand{\arraystretch}{1.05}
\begin{tabular}{lcc}
\toprule
\textbf{Method} & \textbf{Latency (ms)}\\
\midrule
Naive full model in TEE &  1792.42 \\
Model Partitioning \cite{mo2020darknetz,sun2020shadownet,lee2019occlumency,tramer2018slalom}              & 1463.80   \\
Operation-aware Execution \cite{hou2021model,nayan2025secureinfer}       & 3127.73     \\
\rowcolor{gray!20}
\ourtitle            & 1082.67\\ 
\bottomrule
\end{tabular}
\vspace{-3mm}
\end{wraptable}
\textit{Comparison with SOTA TEE-based frameworks.} Existing TEE-based frameworks generally fall into two categories: model partitioning \cite{mo2020darknetz,sun2020shadownet,lee2019occlumency,tramer2018slalom}, where only part of the model is executed inside the enclave while the rest runs in an untrusted region, and operation splitting, where individual computations are offloaded across trusted and untrusted domains. As shown in Table~\ref{tab:tee_comparison}, both approaches introduce a noticeable latency gap compared to GPU-based inference due to frequent enclave transitions and communication overhead.

In contrast, \ourtitle achieves comparable and in some settings even lower inference latency than GPU execution, while still ensuring full-model execution entirely within the TEE, providing far superior security benefits highlighted in C4.

\begin{wraptable}{r}{0.45\textwidth}
\centering
\vspace{-15pt}
\caption{\ourtitle and acceleration methods on DeiT-small on ImageNet.}
\scriptsize
\label{tab:compress_comparison}
\setlength{\tabcolsep}{5pt}
\renewcommand{\arraystretch}{1.05}
\begin{tabular}{lcc}
\toprule
\textbf{Model} & \textbf{Acc.(\%)} & \shortstack{\textbf{TEE Ope.}\\\textbf{Support}} \\
\midrule
Baseline (84.1 MB) & 79.72 & -- \\
Pruning~\cite{yu2022width}  (33 MB)              & 60.51    & $\times$ \\
2-bit Quantization~\cite{li2022q}  (6.18 MB)  & 71.90   & $\times$ \\
\rowcolor{gray!20}
\ourtitle    (4.63 MB)        & \imsixtysix & \checkmark \\
\bottomrule
\end{tabular}
\vspace{-15pt}
\end{wraptable}
\textit{Comparing with traditional acceleration techniques.} historically, model acceleration techniques such as low-bit quantization~\cite{li2022q} and pruning~\cite{yu2022width} have been widely used to reduce computational cost. As shown in Table~\ref{tab:compress_comparison}, both approaches significantly degrade model accuracy compared to the original model.

Moreover, these software optimizations lack hardware support in TEE environments, particularly for low-bit (2-bit) quantization and aggressive pruning configurations. In contrast, as highlighted in Table~\ref{tab:compress_comparison}, \ourtitle is the only approach that can reduce model computation while maintaining reasonable performance on ImageNet, with a lower memory footprint.

\begin{wraptable}{r}{0.36\textwidth}
\centering
\vspace{-12pt}
\caption{White-box (WB) and black-box (BB) attacks on DeiT-B. Attack success rate (ASR, \%) is reported.}
\label{tab:adv_bb_wb_comparison}
\scriptsize
\setlength{\tabcolsep}{4pt}
\renewcommand{\arraystretch}{0.95}
\begin{tabular}{lcc}
\toprule
\textbf{Type} & \textbf{Setting} & \textbf{ASR (\%)} \\
\midrule
\multirow{2}{*}{Adversarial Attacks}
& WB~\cite{NEURIPS2024_24f8dd1b} & 99.40 \\
& BB~\cite{10.1016/j.asoc.2025.113686} & 60.00 \\
\midrule
\multirow{2}{*}{Bit-Flip Attack}
& WB~\cite{rakin2021t} & 100.0 \\
& BB & 0.0 \\
\bottomrule
\end{tabular}
\end{wraptable}
\textbf{Security benefit of hosting the entire model on TEE (\textit{Answers for C4}).} Beyond accelerating vision model execution within TEEs, \ourtitle also shifts the deployment closer to a black-box attack spectrum, as illustrated in Figure~\ref{fig:overview}. To evaluate this transition, we analyze the model’s robustness under black-box attack scenarios. Table~\ref{tab:adv_bb_wb_comparison} highlights the vulnerability of vision models under both white-box and black-box threat settings. The results show that attack success rates are significantly higher in the white-box setting, while black-box attacks are considerably less effective. This gap indicates that reducing model exposure strengthens robustness against adversaries with limited access. Consequently, \ourtitle not only accelerates inference within TEEs but also helps preserve the security of the accelerated models, which is critical in edge vision applications.

\subsubsection{Ablation Study}
\begin{table}[t!]
\centering
\caption{Accuracy and model size comparison (Original vs Inside TEE) on ImageNet for DeiT-Small under different computation reduction factors $\mathcal{C}$. Accuracy (\%) is reported.}
\label{tab:imagenet_computation_budget_accuracy}
\scriptsize
\setlength{\tabcolsep}{6pt}
\renewcommand{\arraystretch}{0.95}
\begin{tabular}{lcccccc}
\toprule
\multirow{2}{*}{\textbf{Model}} 
& \multicolumn{2}{c}{\textbf{Original}} 
& \multicolumn{2}{c}{$\boldsymbol{\mathcal{C}=0.66}$} 
& \multicolumn{2}{c}{$\boldsymbol{\mathcal{C}=0.5}$} \\
\cmidrule(lr){2-3}
\cmidrule(lr){4-5}
\cmidrule(lr){6-7}
& \textbf{Model Size (MB)} & \textbf{Acc.} 
& \textbf{Model Size (MB)} & \textbf{Acc.} 
& \textbf{Model Size (MB)} & \textbf{Acc.} \\
\midrule
DeiT-Small & 84.11 & 79.72 & 4.63 & \imsixtysix & 3.81 & 75.79 \\
\bottomrule
\end{tabular}
\vspace{-10pt}
\end{table}

\noindent\textbf{Different Dataset.} Table~\ref{tab:imagenet_computation_budget_accuracy} reports the performance of DeiT-small on ImageNet using different reduction factors, further demonstrating the effectiveness of \ourtitle on more complex tasks. Despite the aggressive reduction to only 3.81 MB, the reduced DeiT-Small model still achieves competitive accuracy compared to lightweight vision models of similar size, such as ResNet20~\cite{he2016deep} and ShuffleNetV2~\cite{zhang2018shufflenet}, on ImageNet. 

\noindent\textbf{Different Architecture.} Table~\ref{tab:cifar10_computation_budget_accuracy} evaluates DeiT-Base on CIFAR-10 under multiple computation reduction factors. The result is consistent across different datasets as \ourtitle provides a reasonable trade-off between accuracy and model acceleration/size.

\begin{table}[t!]
\centering
\caption{Accuracy and model size comparison (Original vs Inside TEE) on CIFAR-10 for DeiT-Base under different computation reduction factors $\mathcal{C}$. Accuracy (\%) is reported.}
\label{tab:cifar10_computation_budget_accuracy}
\scriptsize
\setlength{\tabcolsep}{6pt}
\renewcommand{\arraystretch}{0.95}
\begin{tabular}{lcccccccc}
\toprule
\multirow{2}{*}{\textbf{Model}} 
& \multicolumn{2}{c}{\textbf{Original}} 
& \multicolumn{2}{c}{$\boldsymbol{\mathcal{C}=0.5}$} 
& \multicolumn{2}{c}{$\boldsymbol{\mathcal{C}=0.4}$} 
& \multicolumn{2}{c}{$\boldsymbol{\mathcal{C}=0.2}$} \\
\cmidrule(lr){2-3}
\cmidrule(lr){4-5}
\cmidrule(lr){6-7}
\cmidrule(lr){8-9}
& \textbf{Model Size (MB)} & \textbf{Acc.} 
& \textbf{Model Size (MB)} & \textbf{Acc.} 
& \textbf{Model Size (MB)} & \textbf{Acc.} 
& \textbf{Model Size (MB)} & \textbf{Acc.} \\
\midrule
DeiT-Base  & 327.33 & 97.55 & 12.83 & 97.22 & 10.47 & 97.04 & 5.80 & 95.57 \\
\bottomrule
\end{tabular}
\vspace{-10pt}
\end{table}

\begin{wraptable}{r}{0.5\textwidth}
\centering
\vspace{-15pt}
\caption{Impact of each component of TEE-X on DeiT-S on the CIFAR-10 dataset.}
\vspace{-2mm}
\scriptsize
\label{tab:component_abla}
\setlength{\tabcolsep}{5pt}
\renewcommand{\arraystretch}{1.05}
\begin{tabular}{lcc}
\toprule
\textbf{Method} & \textbf{Acc.(\%)} &  Inference Time (s)\\
\midrule
Baseline (Naive full model TEE) & 97.29 & 4.43 \\
\ourtitle (w/o modularization)           & 97.29    & 1.80 \\
\ourtitle (w/o vectorization)       & 96.48  & 1.77 \\
\rowcolor{gray!20}
\ourtitle         & 96.48 & 1.08 \\
Host (GPU)         & 97.29 & 1.02 \\
\bottomrule
\end{tabular}
\vspace{-10pt}
\end{wraptable}

\noindent\textbf{Ablation study of each component of \ourtitle.}
We compare the performance impact of each component of the proposed \ourtitle in terms of two primary metrics: accuracy \& latency. We take two baselines: worst-case (naively running the entire model on TEE) and best-case (host GPU). It is evident from Table~\ref{tab:component_abla} that the proposed \ourtitle performance (1.08s) is par with GPU host-only inference, with a substantial gain compared to naive TEE implementation. If we remove the modularization component, then the model accuracy remains unchanged while inference time increases by 80 \% (1.8s) compared to \ourtitle, underscoring the need for the modularization step. Similarly, removing the vectorization component will slow \ourtitle performance by 77 \% (1.77s), emphasizing that both modularization and vectorization have been critical towards achieving on-par GPU acceleration.

\begin{wraptable}{r}{0.43\textwidth}
\centering
\vspace{-20pt}
\caption{ Effect of Different Computation Budget Assignment Strategy in TEE for Different $\mathcal{C}$.}
\label{tab:ablation_uniform_vs_ours}
\scriptsize
\setlength{\tabcolsep}{6pt}
\renewcommand{\arraystretch}{0.95}
\begin{tabular}{lcc}
\toprule
\textbf{Method} & $\boldsymbol{\mathcal{C}=0.2}$ & $\boldsymbol{\mathcal{C}=0.4}$ \\
\midrule
Uniform Budget & 91.27 & 93.55 \\
\rowcolor{gray!20}
\ourtitle (Ours)    & 94.64 & 96.40 \\
\bottomrule
\end{tabular}
\end{wraptable}
\noindent\textbf{Ablation of Computation Budget Assignment Strategy.} Table~\ref{tab:ablation_uniform_vs_ours} shows the effect of different 
computation budget assignment strategy on the accuracy performance of a model. An alternative strategy would be to uniformly assign the same computation budget to all layers. The results clearly indicate that the proposed \ourtitle following Eqn.~\ref{eq:rank_assignment} to assign computation budget across different modules helps maintain model performance, especially under strict budgets.

\section{Conclusion}
In this work, we introduce ~\ourtitle, a TEE-aware acceleration framework that enables efficient inference of large vision models in edge applications within secure memory. \ourtitle addresses the bottlenecks to the full deployment of ViTs in TEEs: limited secure memory, high latency, and the constraint of preserving accuracy under a strict computation budget. To achieve this, we propose sensitivity-based modularization, TEE-aware computation-budget allocation, and SIMD-based vectorized execution within Arm TrustZone. Overall, \ourtitle demonstrates a practical way to achieve near-GPU inference latency with high accuracy while keeping the entire model execution within trusted memory, thereby avoiding the security limitations of partitioning- or offloading-based approaches.

\bibliographystyle{plain}
\bibliography{ref}

\end{document}